# Knowledge Graphs for Innovation Ecosystems

Alberto Tejero[1], Víctor Rodríguez-Doncel[2] and Iván Pau[3]

[1] Center for Technology Innovation (CAIT), Universidad Politécnica de Madrid, alberto.tejero@upm.es

[2] Ontology Engineering Group (OEG), Universidad Politécnica de Madrid, vrodriguez@fi.upm.es

[3] Departamento de Ingeniería Telemática y Electrónica (DTE), Universidad Politécnica de Madrid, ivan.pau@upm.es

**Abstract:** Innovation ecosystems can be naturally described as a collection of networked entities, such as experts, institutions, projects, technologies and products. Representing in a machine-readable form these entities and their relations is not entirely attainable, due to the existence of abstract concepts such as 'knowledge' and due to the confidential, non-public nature of this information, but even its partial depiction is of strong interest. The representation of innovation ecosystems incarnated as knowledge graphs would enable the generation of reports with new insights, the execution of advanced data analysis tasks. An ontology to capture the essential entities and relations is presented, as well as the description of data sources, which can be used to populate innovation knowledge graphs. Finally, the application case of the Universidad Politécnica de Madrid is presented, as well as an insight of future applications.

## 1. INTRODUCTION

Innovation is presented as one of the most important tools for improving the competitiveness of organizations [1-3]. Innovation is a process composed of a set of technical, organizational, productive and commercial stages that lead to the successful launch in the market of new manufactured products or new services [4].

Eurostat, the statistical office of the European Union, defines innovation as "*the use of new ideas, products or methods where they have not been used before*"; in other words, innovation is "*the application of knowledge to produce new knowledge*" [5]. The generation, transfer and application of knowledge has been studied in open systems with some logical meaning, in the so-called innovation ecosystems. Innovation ecosystems are "*dynamic systems made up by actors and connected by knowledge flows based on the relationships of these actors*" [6]. Innovation ecosystems are important because they host the most valuable future assets in the information society, and they determine the competitiveness and economic growth of the regions where they are hosted.

Whereas the traditional study of innovation ecosystems has been of a qualitative nature, recent technological advances and the massive availability of open data sources related to innovation make quantitative analysis also possible [7]. Upon a due numeric description of innovation ecosystems, system analysis can be applied, and innovation ecosystems, previously seen as black boxes characterized by input and output resources, can be described now by a number of numeric parameters. In addition, if different time-related states can be observed in innovation ecosystems, process mining techniques could lead to an unprecedented understanding of the achievement of innovation ecosystems. Interactive process mining techniques, a new approach in process mining techniques, would enable the understanding of the states that suppose a bottleneck in the appearance of a new product, or what are the real interactions between the different agents of the ecosystem for the appearance of a certain innovation [8,9,10].

In the meantime, the construction of Knowledge Graphs for every conceivable domain has been a trend in the last few years. With respect to innovation knowledge graphs, a number of resources of interest has been made available, such as the linked data public recently published by the European Patent Office, the Springer Nature SciGraph for the scholarly domain, or Thomson Reuters' Permid database on organizations (now Refinitiv's). Whereas these resources are the primary source of information to populate an innovation knowledge graph, their inception was made with other purposes and specific data models and ontologies for innovation models graphs are necessary.

Innovation ecosystems have a relevant number of entities with very rich interactions among them. As other domains in engineering, it is possible to apply a multilayer approach to address systematically the inherent complexity. Each layer corresponds to a view of the full knowledge graph capturing different aspects in the relationship among the entities involved.

The main contributions of this paper are (i) establishment of a framework for the development of vertically scalable knowledge graphs for innovation ecosystems; (ii) the definition of a data model, based on ontologies, for the base definition of the innovation ecosystem graph; (iii) and the application of the innovation knowledge graph to the Universidad Politécnica de Madrid (UPM) case study.

This paper is organized as follows: Section two shows the description of the framework for innovation ecosystems, Section three is dedicated to the presentation of knowledge graphs for innovation ecosystems, Section four shows the enabled studies of innovation ecosystems, in Section five the related work in this area of knowledge and finally, in Section six the conclusions.

## 2. DESCRIPTION FRAMEWORK FOR INNOVATION ECOSYSTEMS

### 2.1 INTEGRATED INNOVATION MANAGEMENT SYSTEM

Innovation managers have several tools to describe and stimulate innovation ecosystems. The use of these tools ranges from supporting to business development to creating events that favor certain types of interactions that result in final products. However, deciding the most efficient actions for an innovation ecosystem is not easy. Knowing the dynamics of the ecosystem to identify the catalyzing elements, as well as predicting the effectiveness of certain measures are complex tasks that offer few guarantees a priori.

An integrated innovation management system should facilitate decision-making by innovation managers. To do this, they must adequately model the ecosystem and apply analysis techniques that offer valid indicators to the managers. However, these systems are difficult to implement given the complexity of innovation ecosystems.

This section presents a framework for incremental modeling and analysis of innovation ecosystems. The framework will enable the creation of management tools that support advanced decision-making in innovation ecosystems.

### 2.2.1 GRAPH-BASED MODELING

The modeling of the internals of an innovation ecosystem involves the capture of the entities involved, the artifacts generated or acquired and the different interactions between entities and artifacts. Graphs are suitable mathematical structures for modeling both entities and artifacts as nodes, and the interactions between those nodes as arrows. In addition, once the modeling is done, graphs allow the use of mature analysis methods, as Social Network Analysis, which allow for characterize the behavior quantitatively, answer complex questions about its current state, find bottlenecks or optimal paths in processes and even predict how certain actions may affect the general state of the system.

## 2.2 ONTOLOGIES TO SUPPORT THE FORMALIZATION

To take advantage of the power of graphs as modeling elements of innovation ecosystems, a correct conceptualization of entities, artifacts and interactions is necessary. In addition, this conceptualization must be formalized so that it can be used as an engineering tool that allows not only the construction of knowledge graphs but also the incremental construction of increasingly complete systems.

Ontologies facilitate both the conceptualization process and its formalization, enabling sophisticated engineering methods for its treatment. In addition, their associated technologies facilitate the creation of knowledge graphs.

In the system proposed in this paper, ontologies guide the design of the knowledge graph for innovation ecosystems, establishing both the entities and the interactions that must be captured. The information that is not properly formalized in the ontologies cannot be part of the knowledge graph. This guarantees both the semantic coherence of the graph and its ability to link and interoperate with other knowledge graphs related to innovation ecosystems.

## 2.3 LAYERED MODELING

Although the use of ontologies eases the creation of an efficient and world-linked innovation ecosystems knowledge graph, the complete development also requires the definition of a proper method to take on it systematically. Layered architectures are well-known patterns in engineering to address systematically the modeling of systems with a high number of interactions. Multilayer representation is designed so that each level adds only the relationships between entities necessary to satisfy their semantic requirements. In this way, it is possible to approach the problem in a gradual manner.

In a multilayer system, the base layer should allow both the representation of basic and most tangible interactions among entities and the addition of new layers without functionality restriction. In the case of the proposed framework for innovation ecosystems, the base layer, called INNEO and presented later, includes common entities and artifacts officially registered in most of the innovation ecosystems such as research centers, startups, articles, etc. The refinement of the graph with more subtle interactions and intangible or uncommon entities will be defined in other layers and are out of the scope of this paper.

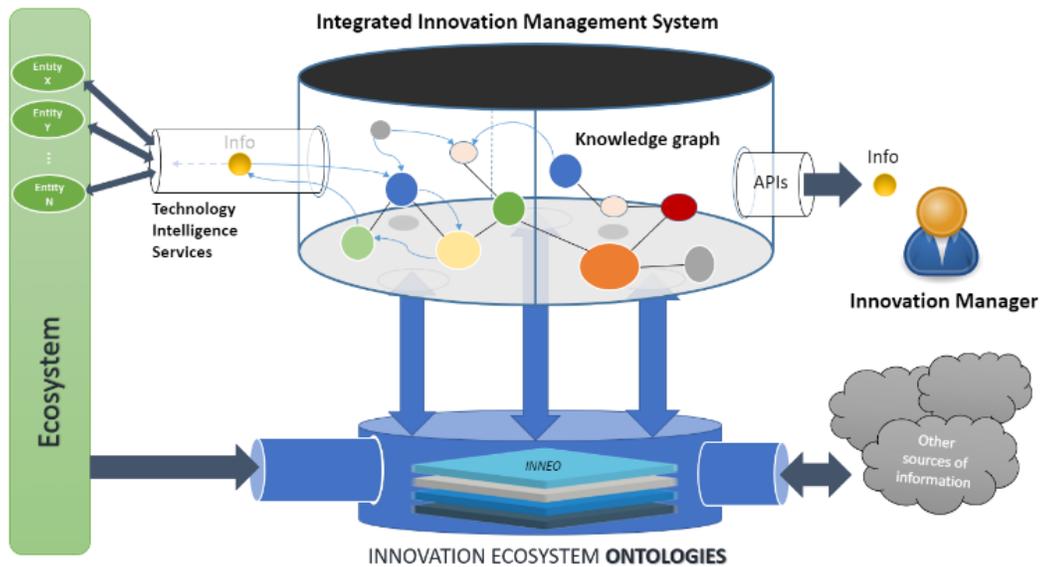

**Figure 1**. Integrated Innovation Management System (Own elaboration)

Figure 1 shows the final high-level architecture of the Integrated Innovation Management System based on the proposed framework. The management system collects information from the innovation ecosystem and outputs patterns and indicators to improve the decisions of the innovation manager. The basis for obtaining the patterns and indicators is the analysis of the generated knowledge graph. The knowledge graph is generated from the information collected by the different entities of the innovation ecosystem. This information, before forming part of the graph, must be formalized through the different ontologies defined in the system, including the INNEO ontology.

## 3. KNOWLEDGE GRAPHS FOR INNOVATION ECOSYSTEMS

Knowledge graphs are no other thing than structured information, where some entities are represented as nodes, their attributes as node labels and the relationship between entities are represented as edges. Knowledge graphs have become an essential asset for many companies in the information society, and we would not conceive Google, Facebook or LinkedIn without their respective knowledge graphs. Knowledge graphs turn data into knowledge, either if they are general (such as Freebase's [11]), focused on the linguistic domain, such as WordNet [12] or more domain specific, such as the Springer SciGraph on research. Knowledge graphs have become important resources for many AI and NLP applications such as information search, data integration, data analytics, question answering or context-dependent recommendations.

Whereas there is no specific technology behind the idea of knowledge graph, one of its most convenient representations is the one adhering to the specifications published by the W3C

Consortium in the area of Semantic Web: data represented in RDF, supported by OWL ontologies acting as data models and a clean publication using the Linked Data approach (e.g. dereferenceable URIs, content negotiation, etc.).

### 3.1 THE INNOVATION ECOSYSTEM ONTOLOGY

On despite of these efforts, to date, no knowledge graph specifically targeting innovation ecosystems has been described, nor the data model behind it. In order to cover this gap, we present here the *INNovation Ecosystem Ontology* (INNEO), which is an ontology of innovation ecosystems, which can serve as data model for the key information of these ecosystems. The goal of this effort is to support the decision making of managers, actors and policy makers of the innovation ecosystem. The INNEO ontology can be the cornerstone of an architecture comprising different technologies and data sources, becoming the reference data model on which, the interoperation of different information sources pivot.

More specifically, the INNEO ontology can be used for (i) the representation of innovation-related information (such as patents, universities or researchers) in an ecosystem at a certain moment; (ii) the conceptualization of innovation ecosystems as dynamic graphs, where the temporal evolution of the innovation ecosystem components is captured and (iii) as a data model of information to be used in future data analysis, possibly including event mining and simulation.

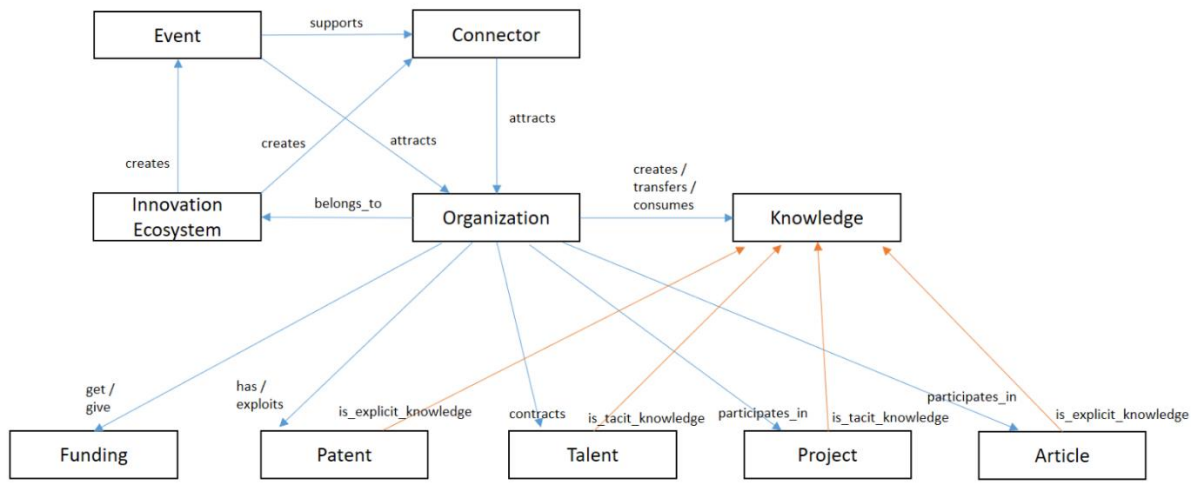

**Figure 2.** Main classes and properties of the INNEO ontology (Own elaboration)

The INNEO ontology, whose URI is https://github.com/atejerolopez/innovation_knowledge_graph, is an OWL ontology and it contains 10 classes: *InnovationEcosystem, Knowledge, Organization, Event, Connector, Funding, Patent, Talent, Project, Article*. Figure 2 represents these classes, along with the most relevant properties. The central class is Innovation Ecosystem, which had been never modelled before –to the best knowledge of the authors. Innovation ecosystems comprise different Organizations, which create Events and Connectors [7] –entities in turn attracting organizations and putting them in touch. The class Knowledge represent an abstract concept, which cannot be directly perceived, quantified or represented in many cases, such as tacit knowledge. The rest of the classes represent concepts which already have (or can easily have) a description, and they are probably entries in existing databases –think of Article, Patent or Project.

We contend that Knowledge, as an abstraction, is manifested in patents, research articles, projects, talent and other derived evidences. We cannot measure Knowledge, but its manifestations. Besides the object properties relating these classes (using simple name such as belongsTo, etc.), only a few datatype properties have been defined –instead, each of the instantiable classes, has been mapped to other terms defined by external, well-known vocabularies and ontologies.

Defining a metrics of knowledge, or formulas and indicators to quantify knowledge, is not an unconventional practice. The transfer of knowledge in university ecosystems has been numerically evaluated [13], and knowledge at organization level has also been quantified –for example Matoskova [14] has combined traditional methodologies and indicators such as Knowledge Capital Earnings, Human Capital Index, Knowledge Management Scan, the Skandia Navigator, Citation-Weighted Patents or simply opinion-based surveys. Beyond these conventional methods, the existence of vast, new, open data sources justifies a new approach with advantages such as its easy calculation, its dynamic nature and the reliability of its sources.

The INNEO ontology has been designed following the Neon methodology for the design of ontologies [15], and a formal design process was followed. In particular, a list of competency questions was drafted by different experts in innovation. Among the most relevant questions, we could mention the following:

- What are the organizations working in the area of knowledge X at scientific level?
- Does the organization X share patents with the organization Y?
- In which technologies is patenting the organization X?
- Which organizations have received funding in year X?
- What are the organizations that hire staff of the area of knowledge X?
- How can collaborate the organization X with other organizations in the area of knowledge X?

The INNEO ontology is publicly licensed (as CC-BY) and has passed the quality tests and verifications suggested by the Oops pitfall detector system [16].

### 3.2 THE INNOVATION KNOWLEDGE GRAPH

An innovation knowledge graph is a set of informational items on innovation, a collection of innovation-related entities and their relations. In one way or another, every large organization with interest in innovation has maintained databases on innovation. For example, Universidad Politécnica de Madrid uses the S2i UPM system, which permits access to internal information of the university, related to its organizational structure, human resources and exploitation results, among others. S2i is fed from internal data sources and does not leak information, which is in part confidential because of its strategic nature and is in part personal (human resources information).

However, the business game is changing and a number of open data sources have sprouted to populate the knowledge graph. Some of them have the two most desired features: they are open data sources (ready to be reused) and they are structured as linked data, making the integration tasks almost immediate. Table 1 identifies a number of data sources to populate the graph.

| Domain | Publishers | Data Sources | Main entities in the graphs |
|---|---|---|---|
| Patents | European Patent Office (EPO), Korean Intellectual Property Office (KIPO) | EPO patents, KIPO Korean patent office | Patents, Inventors (people), Applicants (organization), References to papers, Patent classifications (knowledge domains). |
| Article, talent, funding | Springer Nature, Trier University, Microsoft Academic Graph | Springer Nature SciGraph, DBLP, Semantic Scholar | Research papers, authors (persons, institutions), topics, grants, funding projects. |
| Projects | EU Publications Office | CORDIS, TED Portal | Public tendering, companies, sectors. |
| Organizations | Thomson Reuters, euBusinessGraph, OpenCorporates, Crunchbase | Permid, euBusinessGraph | Organizations, people (talent), financial instrument (funding) |

**Table 1**. Some well-known data sources that could be used to populate the graph (Own elaboration)

Patents are available as linked data only from recent times. In April 2018, the European Patent Office launched a product-level solution, dumping patent information as linked data with weekly updates [22]. Previously, the Korean Patent office had made a similar effort, publishing Korean and foreign patents as linked data. In addition, non-official transformations have been made available, such as Hassan's translation of the US patent office information [21].

Information on organizations is also abundant. PermID is an open database on several millions of organizations and related information, such as the key persons, financial instruments and quotes; the H2020 EU project euBusinessGraph has embarked on a similar endeavour. Larger databases on companies are also available under payment modalities, such as OpenCorporates or Crunchbase, which provide company information such as founding information, acquisitions, investors, and other related people. Finally, official registry information on companies is of public nature per se, and the official registries have become a source of massive data about companies –e.g. see the UK Companies House. A complete study on the openness of official registries is maintained by the Open Knowledge Foundation[1], revealing that more and more registries are publishing information with an open license and in a machine-readable form, such as Argentina, Australia, Canada, France, United Kingdom, Japan, Norway, Russia, Singapore, Taiwan, Thailand or Ukraine.

Again, official categorizations for industrial and economic activities exist, but not many have been ported to RDF. For example, one of the most popular ones, NACE, the statistical classification of economic activities in the European Community, is published as XML but only ported to RDF by non-official initiatives (e.g. such as the H2020 EU-funded OpenBudgets project). A list of different categorizations is shown in Table 2.

---
[1] https://index.okfn.org/dataset/companies/

| Name | Scope | Link |
|---|---|---|
| UN ISIC Rev 4 | Global | http://unstats.un.org/unsd/cr/registry/isic-4.asp |
| European Community NACE Rev 2 | European Union | http://epp.eurostat.ec.europa.eu/portal/page/portal/nace_rev2/introduction |
| North American Industry Classification System 2017 | United States | https://www.census.gov/cgi-bin/sssd/naics/naicsrch?chart=2017 |
| UK SIC Classification 2007 | United Kingdom | http://www.ons.gov.uk/ons/guide-method/classifications/current-standard-classifications/standard-industrial-classification/index.html |
| Russian Classification of Economic Activities | Russia | http://www.gks.ru/metod/classifiers.html |
| Australian and New Zealand Standard Industrial Classification (ANZSIC) 2006 | Australia New Zeal. | http://www.stats.govt.nz/methods/classifications-and-standards/classification-related-stats-standards/industrial-classification.aspx |
| India National Industrial Classification 2004 | India | http://www.mca.gov.in/MCA21/dca/efiling/NIC-2004_detail_19jan2009.pdf |
| Nomenclature d'activités française (2008) | France | http://www.insee.fr/fr/methodes/default.asp?page=nomenclatures/naf2008/naf2008.htm |
| Dansk Branchekode 2007 | Denmark | http://www.dst.dk/en/Statistik/dokumentation/Nomenklaturer/DB.aspx |
| Norway Standard Industrial Classification | Norway | http://stabas.ssb.no/ItemsFrames.asp?ID=8118001&Language=en&VersionLevel=classversion&MenuChoice=Language |
| NACELUX Rev 2 | Luxemburg | http://www.environnement.public.lu/dechets/informations_pratiques/code_nace.pdf |
| Finland TOL 2008 | Finnland | http://tilastokeskus.fi/meta/luokitukset/toimiala/001-2002/kuvaus_en.html |
| Singapore Standard Industrial Classification | Singapre | http://www.singstat.gov.sg/methodologies-standards/statistical-standards-and-classifications/SSIC |
| Bulgarian Classification of Economic Activities | Bulgaria | http://www.nsi.bg/bg/taxonomy/term/тема-статистически-класификации |
| Viet Nam Standard Industrial Classification | Vietnam | http://www.gso.gov.vn/default.aspx?tabid=728 |
| Thailand Standard Industrial Classification | Thailand | http://riped.utcc.ac.th/wp-content/uploads/2016/03/LFS-Industry-Code-2013.pdf |
| NACE-BEL 2008 | Belgium | http://statbel.fgov.be/nl/statistieken/gegevensinzameling/nomenclaturen/nacebel/ |

**Table 2.** Official categorizations for industrial and economic activities that could be used to populate the graph (Own elaboration)

The scholar domain is profuse in linked data initiatives. Springer Nature SciGraph is a linked open data platform for the scholarly domain, which not only includes research papers but also persons, grants, product market codes or a version in RDF of the Global Research Identifiers Database (which contains information on about 100,000 institutions around the globe). Many other datasets on research articles (or their metadata) are available, such as Semantic Scholar, with 45 million published research papers in Computer Science, Neuroscience, and Biomedical sciences or DBLP[2], providing service (active for the last 25 years) offering bibliographical information as RDF. The Microsoft Academic Graph, available through the Open Academy Society, contains an impressive amount of 170M papers online. These are not the only initiatives, and many others have sprouted, like the IOS Press linked data[3] or the Ace Knowledge Graph [24].

There is no open dataset on talent of relevance, but a relatively reliable collection can be built from the academic and patent-related datasets.

Projects and funding have also several devoted datasets. Information on the EU research projects tracked by the Cordis[4] is available for download at the European Data Portal, which also includes datasets on public tendering. The TED (Tenders Electronic Daily) "*publishes 520 thousand procurement notices a year, including 210 thousand calls for tenders which are worth approximately €420 billion*"[5] –a fraction of this information has been also published and enriched by the H2020 project TheyBuyForYou.

---

[2] http://dblp.org
[3] http://ld.iospress.nl
[4] https://cordis.europa.eu

An initiative that deserves mention, because of its relationship with the approach presented in this paper and with university-type ecosystems, is the Hercules project[6]. This initiative, promoted by the CRUE Spanish Universities, aims to create a new model of collaboration between universities for the development of their management systems. Specifically, the project aims to create a semantic architecture for the research management system of universities. In short, joint exploitation of research information of all universities: knowledge of scientific production, ease of technology transfer, information to improve the mobility of teachers and researchers, etc.

### 3.3 INNOVATION KNOWLEDGE GRAPHS APPLICATIONS

#### 3.3.1 ENRICHMENT OF PATENTS MANAGEMENT USING external INFORMATION SOURCES

The potential of the Innovation Knowledge Graph is very large; it can already be applied at different levels. At a particular level, in the case of an innovation ecosystem of university type, it allows to enrich the information that is usually available with the incorporation of external sources to those of the university itself. This fact enables the obtaining of context knowledge, which helps in an important way in the usual process of decision-making. In the case of the UPM, the connection of information obtained from the UPM S2i database, for example, the patents generated from the university, with contextual information such as scientific articles, projects and information from organizations, allows obtaining knowledge maps such as the one shown in Figure 3.

---

[5] https://ted.europa.eu
[6] http://tic.crue.org/hercules/

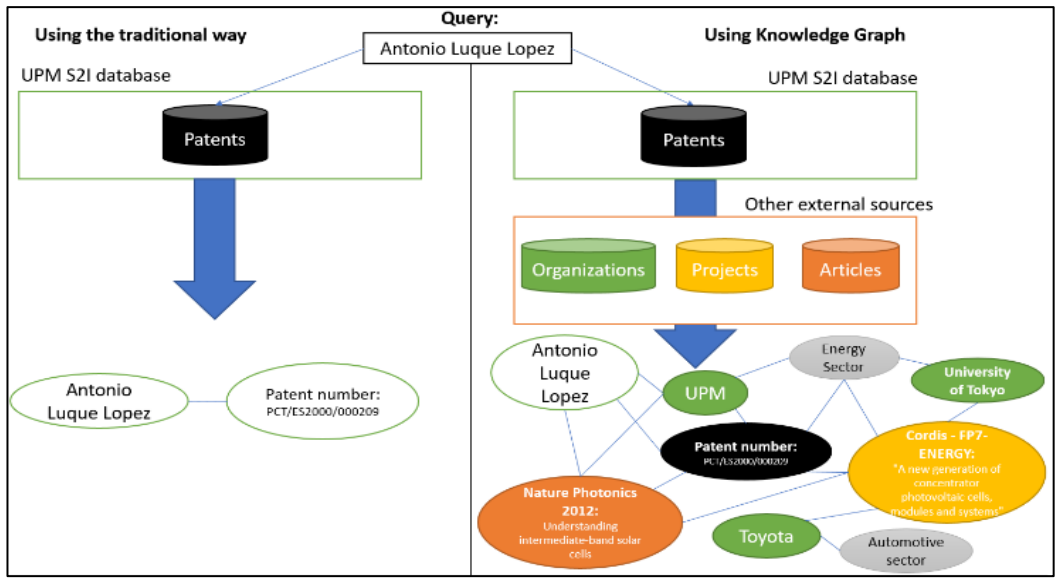

**Figure 3**. Application case of the Innovation Knowledge Graph with the Universidad Politécnica de Madrid S2i database (Own elaboration)

Thanks to this new structure, more complex questions can also be resolved. For example, in the case of the knowledge graph generated for the UPM, shown in Figure 3, to the question 'Does the UPM researcher, Antonio Luque Lopez, have a relationship with the automotive sector?', the system response would be: Yes, also showing that the relationship was established through the company 'Toyota' within the European project 'A new generation of concentrator photovoltaic cells, modules and systems', in relation to the patented technology through the number of international application 'PCT/ES2000/000209', of which the article 'Understanding intermediate-band solar cells' has also been written in the journal 'Nature Photonics', etc. Figure 4 shows the High-level view of the answer generation system.

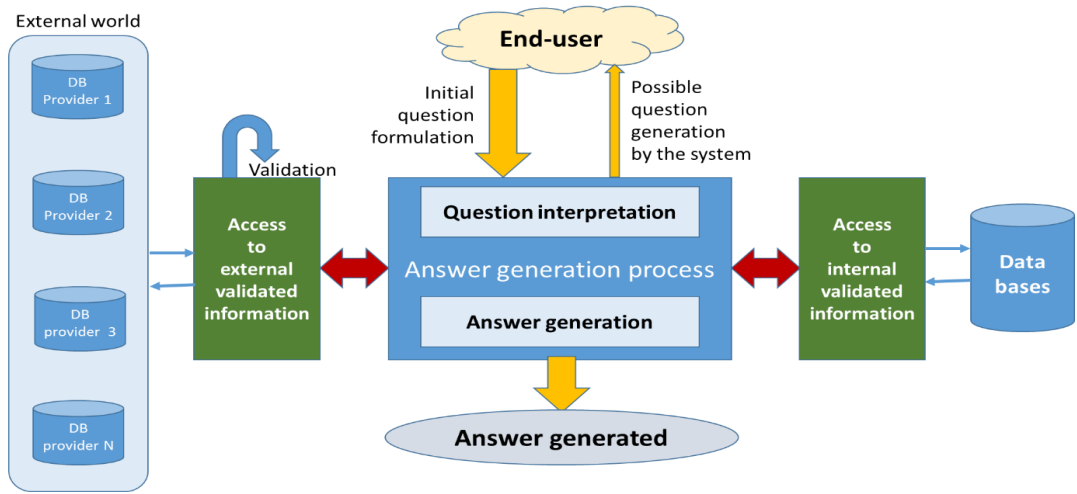

**Figure 4.** High-level view of the answer generation system (Own elaboration)

### 3.3.2 FUTURE APPLICATIONS: DIGITAL INNOVATION HUBS

At another scale, the use of the Innovation Knowledge Graph in other collaboration contexts is also possible. For example, the Digital Innovation Hub (DIH), an initiative launched by the European Commission on April 19, 2016 and currently has over 200 fully operational hubs across the EU. A DIH is a regional multi-partner cooperation around specific technical competences (e.g. artificial intelligence and cognitive systems, Internet of Things, etc.), which include organizations like universities, industry, regional development agencies, etc.).

The sources of information typically associated with a DIH are those shown in the example in Figure 5. In this case, as can be seen, they coincide almost entirely with the information sources represented by classes in the INNEO ontology (see Figure 2).

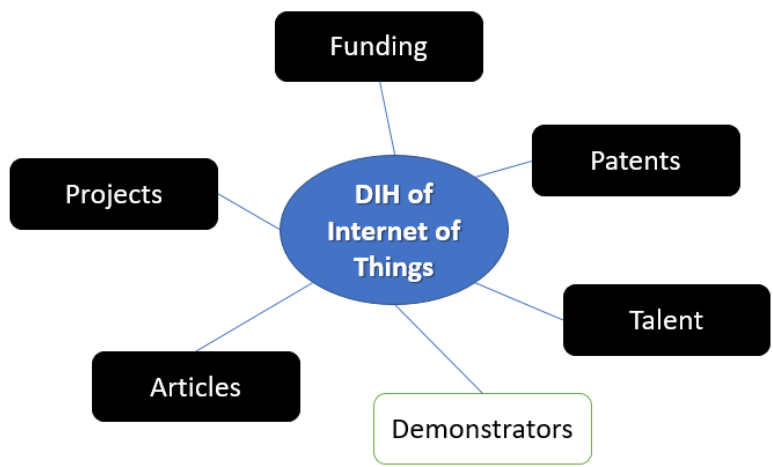

**Figure 5.** Information sources associated to a Digital Innovation Hub of Internet of Things (Own elaboration)

It is important to highlight in this point, that the application of the Innovation Knowledge Graph does not have to be restricted to a single DIH, but to an entire DIH network in relation to the same technical competences or even encompassing the total DIH network. Again, the benefits of combining the information of several DIH or of all, are the multiple possibilities that are open for decision-making with the enrichment and connection of the knowledge of these units of collaboration.

### 4. ENABLED STUDIES OF INNOVATION ECOSYSTEMS

Innovation ecosystems are complex systems made up of actors of different types that interrelate to generate innovation. That is why any tool, approach or method that allows a better integration of the 'different pieces' that make up the 'innovation ecosystem puzzle', is presumed to be of vital importance. The approach presented in this article constitutes a step forward in this regard. Through the use of ontologies and knowledge graph, it is possible to improve the integration of its actors, which benefits not only each one of them, but also the ecosystem, seen as a whole. From the greater integration of the actors with the ecosystem, it is possible to obtain new metrics, as well as generate new analysis processes that result in a better decision-making for their sustainability.

On the other hand, the relevance of innovation ecosystems for the economic development of the regions is something that still, at present, throws some uncertainty. In fact, the performance of innovation, even knowing its multiple and varied benefits, from a purely empirical point of view, is very difficult to demonstrate. Precisely, this is where the true importance of the advances obtained thanks to the development of the ontology and knowledge graph of innovation ecosystems lies. With this new approach, it is possible to create new 'windows' that allow us to see what really happens in the black box of the innovation process.

It is presumed that being able to observe what is really happening in the process, as well as its analysis through tools such as process mining, could help to understand not only why certain steps and intermediate events occur, but also how and even when. This, without a doubt, enables many new possibilities for study.

From the point of view of the tools presented in this article, the framework and the base ontology define an environment for the incremental development of new models of representation. Its modular approach allows different subsequent investigations to capture, on the basis of previous models, other nuances of the relationship of the different agents of the innovation ecosystem. In addition is possible the development of ontologies, later transformed in new information for the knowledge graph, to support relationships specific to the industrial or academic domain where the innovation ecosystem is integrated.

## 5. RELATED WORK

Although there are no studies related to ontologies and knowledge graphs of innovation ecosystems [17], there are works related to ontologies developed for digital ecosystems [18], as well as ontologies generated for some of the sources normally present in the innovation process and, therefore, that are part of the definition of carried out in the present work. Specifically, there are coming up several initiatives to create frameworks and connections with the articles datasets and other sources such as grants, researchers, etc., such as the 'Research Graph dataset for connecting research data repositories using RD-Switchboard' proposed in the paper of Aryani [19]; a patent system ontology for facilitating retrieval of patent related information is proposed by Taduri [20]; or the 'Funding, Research Administration and Projects Ontology' (FRAPO). From a design point, some of these initiatives could be useful for reuse in future iterations aimed at improving the ontology presented in Figure 2.

However, the main themes on which this work is based on theory, that is, literary trends related to ecosystems of innovation, knowledge management and ecosystem management, have been widely analyzed, as shown below.

## 5.1 INNOVATION ECOSYSTEMS

"*There is a book always open to all eyes: Nature*" — Jean-Jacques Rousseau

According to Oxford dictionaries, 'ecosystem' can be defined as follows:

- (Ecology) "*A biological community of interacting organisms and their physical environment*".
- (In general use) "*A complex network or interconnected system*".

The term ecosystem was proposed by the British botanist Arthur Tansley in 1935, to define the basic functional unit in ecology, which encompasses the biotic communities and the abiotic environment of a specific region, where each of them influences the properties of the other. In the last decade, the literature related to innovation reflects a significant growth of interest in the term ecosystem of innovation, as a conceptual mechanism for improving the weakness of structural innovation [25].

The concept of 'innovation system' was an important milestone towards the clarification of innovation processes since the eighties. However, in the found definitions of the innovation system until 1994, little attention was given to details related to the interactions produced between the different actors normally involved in these systems, that is, to the dynamic relationships, focusing more on the innovation process or system in general.

With the concept of ecosystem is when it starts to place more emphasis on the activities and interactions of its actors, a more directed approach to understanding the dynamics of systems and their sustainability [26]. The concept of ecosystem is distinguished from that of system by the use of analogies between ecosystems of ecological-biological type and those of socio-economic type. In fact, an organization cannot be considered in isolation, since it is usually immersed in a network of interdependencies where a change produced in one part of the system can affect others.

According to Jackson [27], "*a biological ecosystem is a complex set of relationships among the living resources, habitats, and residents of an area, whose functional goal is to maintain an equilibrium sustaining state. In contrast, an innovation ecosystem models the economic rather than the energy dynamics of the complex relationships that are formed between actors or entities whose functional goal is to enable technology development and innovation*". The notion of ecosystems provides an attractive metaphor to describe a range of interactions and inter-linkages between multiple organizations [28].

In the literature on socio-economic ecosystems there are many authors who refer, in one way or another, to the actors usually found in these ecosystems, actors such as large, medium and small companies, educational institutions, institutes of research, public entities, venture capital, etc. [29]. As in ecological ecosystems, actors are structured in different roles and functions.

In these ecosystems, whether industrial, business, entrepreneurship, digital or innovation, knowledge appears as an element that underlies aspects such as learning, people, technologies and culture. In fact, in a society like the current one, based on knowledge, it does not have to surprise that knowledge forms an essential part of the different systems and ecosystems, as it is reflected through concepts such as the well-known knowledge triangle, with which the relationship between the areas of education, research and innovation is highlighted [30].

With the increase in connectivity, specialization and knowledge-based products, it becomes evident that being part of a larger system is not only a competitive advantage, but also a necessity to participate in the creation of value [31]. As a result, organizations demonstrate an increasing interest in organizing in these complex ecosystems, with the adoption of a network-centric strategy that allows them to combine their skills to create products that otherwise could not exist. Innovation ecosystems take advantage of the diversity and autonomy of their stakeholders to obtain potential innovative results, which become the central focus of their activities. This new type of collaboration agreements includes a wide range of actors, who closely resemble the biological communities that interact and depend on each other, evolving and responding to the environment in which they exist.

On the other hand, we must be cautious with the use of the term 'ecosystem'. As some authors point out [32], the notion of innovation ecosystem is an interesting development, based on biomimetic thinking, which injects some useful concepts into the economic development dialogue. However, although it is true that it can lead to new scientific truths and reliable methods for knowledge and economic development, the notion in itself does not constitute or provide such truths or methods. Empirical support or rigorous correspondence rules are needed to support the extraction of conclusions when using this type of analogy.

## 5.2 KNOWLEDGE MANAGEMENT

In the last twenty years, developed countries have moved quickly from a society governed by the accumulation of information to another that could be called knowledge-based society, in which information by itself lacks value if it is not contextualized and analyzed for the benefit of its users. The increase in the importance of knowledge as an economic engine has had great implications for the innovation process and its management, as several studies carried out in this field have shown [33].

Good knowledge management provides support for the creation, transfer and application of knowledge in the organization. However, knowledge management is a complex issue. According to Alavi and Lediner [34]:

*"Knowledge management involves distinct but interdependent processes of knowledge creation, knowledge storage and retrieval, knowledge transfer, and knowledge application. At any point in time, an organization and its members can be involved in multiple knowledge management process chains. As such, knowledge management is not a monolithic but a dynamic and continuous organizational phenomenon. Furthermore, the complexity, resource requirements, and underlying*

*tools and approaches of knowledge management processes vary based on the type, scope, and characteristics of knowledge management processes*".

A strategic knowledge management is related to the processes and infrastructures that organizations use to obtain, create and share knowledge for the generation of new strategies and make decisions. A strategic attitude is necessary to achieve a sustainable competitive advantage. On a practical level, organizations are realizing the importance of managing knowledge to remain competitive and grow. Because of this, companies around the world are beginning to dynamically manage their knowledge and innovation.

In the last two decades, Europe and Canada have experienced a startup revolution. During the last decade, other countries than the United States have successfully developed ecosystems for new technology companies. While it is true that Europe and Canada are almost comparable in numbers of newly created technology companies, there is still a great concern about their growth performance. With the United States creating the vast majority of success stories, the challenge in Europe and Canada has focused on success in the later stages of the business development process. This is the so-called 'scale-up' challenge [35].

In the literature review conducted by Cerchione et al. [36] and Centobelli et al. [37] shows that the vast majority of articles analyzed between 1990 and 2016 are focus on knowledge management in SMEs, while only a few documents are responsible for analyzing knowledge management in networks populated by SMEs. Therefore, a research gap is identified in studies with a focus on the analysis of knowledge management of SME networks and, therefore, within innovation ecosystems.

According to Peter Gray [38], "*knowledge can generate economic value when it is used to solve problems, explore opportunities and make decisions*". Although the role of knowledge management for decision support is well recognized, there is a gap between existing knowledge management theory and practice in real-life decision-making. Knowledge management can contribute to decision-making not only by sharing past experiences, but also by providing knowledge and decision-making structures, inside and outside the domain of the problem. Many authors have highlighted the importance that knowledge management has in decision-making. The positive impact of the knowledge management infrastructure on the quality of decisions and on speed has been demonstrated empirically [39]. Therefore, knowledge and time are seem to be two strongly connected resources, something critical for the competitiveness and survival of organizations.

Innovation depends largely on the availability of knowledge. So much so, that the complexity of innovation has been increased by the amount of knowledge available today by organizations. The innovation process depends on knowledge. In fact, "*innovation, that is the application of knowledge to produce new knowledge*" [40].

At the empirical level, there is evidence that effective knowledge management is developed by organizations with a tendency to develop incremental innovations [41]. In more recent studies, Inkinen et al. [42] provide empirical evidence on how various knowledge management practices influence the performance of innovation. Their results are based on data from surveys collected in

Finland during 2013. In their study, the authors [42] consider that companies are able to support the performance of innovation through the strategic management of the knowledge and competence, as well as compensation practices based on knowledge and information technologies. Their study demonstrates, therefore, the importance of knowledge management for the performance of innovation.

Although universities and research centers are mainly public bodies in many countries, "*the degree of impact of university activities on industrial innovation and the nature of the linkage used depend on the industry concerned, as well as the provision of appropriate policy for knowledge transfer*" [43]. For that reason and because universities are cognitive intensity institutions where the primary function is based on knowledge, knowledge management is one of the main objectives of universities. There is a growing belief that knowledge management in universities helps to build the future of a dynamic learning environment, as well as the development and improvement of activities related to the efficiency of knowledge exchange, and the overall performance of the organization.

### 5.3 MANAGEMENT OF INNOVATION ECOSYSTEMS

As in ecology, in innovation ecosystems, complexity requires an adaptive and incremental management for decision-making where the different actors involved can participate and re-feed for their own benefit and for the entire ecosystem.

The management of strategic suppliers such as universities requires collaborative environments and relationships of trust. Open systems or collaborative environments that, in the case of innovation processes, usually require integration of information systems (for example, to develop complex projects concurrently) and, often, supply chain integration.

Buying pencils is not the same as building buildings or acquiring critical semiconductors. This last case is about 'strategic purchases', where it is often even more advisable to do R&D jointly. This is precisely the case of universities and research centers, as strategic knowledge providers. Universities or research centers do not have sales, product development or delivery processes similar to business processes. Their function is not to solve short-term problems with orientation to the detailed specification (all R&D projects are uncertain by nature) and to minimize economic margins. Its function is, as a strategic knowledge provider, to generate R&D in conjunction with other organizations, with a medium and long-term vision, to prepare the next generation of products. In short, create exclusive capabilities for those organizations with which they collaborate. The research groups work with emerging knowledge, which can be incorporated into the market for several years.

The solution is to patiently create business ecosystems around universities and research centers. Select those organizations with sufficient managerial and technological capabilities. Generate strong bonds of trust. Trace joint technological roadmaps in the medium and long term. Start to conduct strategic research in partnership, learning from each other, progressively modulating research to make it more oriented. Jointly design new generations of products. Integrate industrial doctorates into this logic [44]. Permeabilize the borders and facilitate the 'revolving doors' (doors of entry and exit of company personnel to the university and vice versa). Try to have public policies that help accelerate the process, through favorable taxation, preferential financing and direct aid to organizations that decide to follow this path.

All the aforementioned, requires innovation management, and ultimately knowledge, which allows an adaptive and incremental way to meet all the different milestones needed to achieve the objectives of positioning of universities as strategic knowledge providers, but also as orchestrators of this type of ecosystems.

## 6. CONCLUSIONS

The access and management of current information by organizations that are part of an innovation ecosystem is currently limited. Knowledge graphs are presented as a new way to facilitate access not only to the information required by the ecosystem innovation managers, but also to obtain the enriched information, thus expanding, notably, the benefits obtained in the face of decision-making.

This article shows a new framework of vertical and scalable development of knowledge graphs for innovation ecosystems, based on the use of the ontology created within this framework: INNEO. The application of innovation knowledge graph in the case of the UPM demonstrates its usefulness in university-type innovation ecosystems. In addition, this technological proposal shows its great strength and application in other very different scenarios, such as the Digital Innovation Hubs, where the inclusion of new sources of information, such as those from 'demonstrators' or technology demonstrators, is also possible thanks to the vertical scalability that the new framework allows.

The challenges that arise in the long term are based on the creation of new layers for the extension of the framework. However, the immediate lines of work presented for the continuation of this research, based on the current design and structure shown in this article, focus on the creation of new tools and metrics for the management and analysis of information for part of the users of the ecosystem.